\documentclass[onecolumn,showpacs]{revtex4}

\usepackage{graphicx}% Include figure files
\usepackage{dcolumn}% Align table columns on decimal point
\usepackage{bm}% bold math

\input epsf

\begin{document}

\title{\Large Brans-Dicke Theory in Anisotropic Model with Viscous Fluid}

\author{\bf Shuvendu Chakraborty$^1$\footnote{shuvendu.chakraborty@gmail.com}  and
Ujjal Debnath$^2$\footnote{ujjaldebnath@yahoo.com,
ujjal@iucaa.ernet.in}}

\affiliation{$^1$ Department of Mathematics, Seacom Engineering
College, Sankrail, Howrah-711 302, India.\\ $^2$Department of
Mathematics, Bengal Engineering and Science University, Shibpur,
Howrah-711 103, India.}

\date{\today}

\begin{abstract}
In this paper we have considered an anisotropic space-time model
of the Universe in presence of Brans-Dicke (BD) scalar field
$\phi$, causal viscous fluid and barotropic fluid. We have shown
that irrespective of fluid the causality theory provides late time
acceleration of the Universe. If the deceleration occurs in radial
direction and acceleration occurs in transverse direction then the
anisotropic Universe will accelerate for a particular condition of
the power law representation of the scale factors.
\end{abstract}


\maketitle

\section{\normalsize\bf{ INTRODUCTION}}

The standard cosmological models (SCM) only describe decelerated
Universe models and so cannot reproduce the results coming from
the recent type Ia supernovae observations upto about $z\sim 1$
[1] which in favour of an accelerated current Universe. The recent
extensive search leads to some matter field which gives rise to an
accelerated expansion for the Universe. This type of matter is
called Q-matter. This Q-matter behaves like a cosmological
constant [2] by combining +ve energy density and negative
pressure. This Q-matter is either neglected or unknown responsible
for this present Universe. At the present epoch, a lot of works
have been done to  solve this quintessence problem. Most popular
candidates for Q-matter are formatted  by scalar field having a
potential, generates  a sufficient negative pressure. Furthermore,
observations reveal that this unknown form of matter properly
referred to as the {\it dark energy} which capture almost $70\%$
of the Universe. This is confirmed by the very recent W M A P data
[3]. A large number of possible candidates for this dark energy
component has already been proposed and their behaviour have been
studied extensively [4]. Most of these models fit only to
spatially flat $(k=0)$ Friedmann-Robertson-Walker
model [5], though a few models [6] work for open Universe ($k=-1$) also.\\

Recently, a lot of interest have shown by the researchers in the
Brans-Dicke scalar tensor theory, because of its important
possible role in inflationary scenario [7]. Brans- Dicke (BD)
theory is proved to be very effective regarding the recent study
of cosmic acceleration [8]. BD theory is explained by a scalar
function $\phi$ and a constant coupling constant $\omega$, often
known as the BD parameter. This can be obtained from general
theory of relativity (GR) by letting $\omega \rightarrow \infty$
and $\phi=$ constant [9]. This theory has very effectively solved
the problems of inflation and the early and the late time
behaviour of the Universe. Banerjee and Pavon [8] have shown that
the BD scalar tensor theory can potentially solve the quintessence
problem. The generalized BD theory [10] is an extension of the
original BD theory with a time dependent coupling function
$\omega$. In Generalized BD theory, the BD parameter $\omega$ is a
function of the scalar field $\phi$. Banerjee and Pavon have also
shown that the generalized BD theory gives rise to a decelerating
radiation model  where the big-bang nucleosynthesis scenario is
not adversely affected [8]. Modified BD theory with a
self-interacting potential have also been introduced in this
regard. Bertolami and Martins [11] have used this theory to
present an accelerated Universe for spatially flat model. All
these theories conclude that $\omega$ should have a low negative
value in order to solve the cosmic acceleration problem. This
contradicts the solar system experimental bound $\omega\geq500$.
However Bertolami and Martins [11] have obtained the solution for
accelerated expansion with a potential ${\phi}^{2}$ and large
$|\omega|$. Although they have not considered the positive energy
conditions for the matter and scalar field.\\

Scalar dissipative in cosmology may be treated via the
relativistic theory of bulk viscosity [12, 13]. The causal and
stable thermodynamics of Israel and Stewart provide a
satisfactory replacement of the unstable and non-causal theories
of Eckart and Landau and Lifshitz. If viscosity driven inflation
occurs then this necessarily involves non-linear bulk viscous
pressure [12]. A non-linear generalized  Israel-Stewart theory is
a more satisfactory model of viscosity driven by inflation. The
inflationary Universe scenario can solve some outstanding
problems [14] of standard big-bang cosmology. Ellis and Madsen
[15] have considered an FRW model with a minimally coupled scalar
field along with a potential and perfect fluid in the form of
radiation. For brief but comprehensive reviews, we refer to the
papers by Maartens [12], Hiscock and Lindblom [16] and Lindblom
[17]. In a recent communication, Zimdahl [13] discussed how the
truncated theory is needed  a good limit to the full causal
theory in certain physical situations. Banerjee et al [18] have
obtained the exact solution in BD theory in presence of
causal viscous fluid.\\

\section{\normalsize\bf{Basic Equations and Solutions}}

The Brans-Dicke (BD) theory is described by the action (choosing
$8\pi G=c=1$) given by

\begin{equation}
S=\int d^{4} x \sqrt{-g}\left[\phi R- \frac{\omega}{\phi}
{\phi}^{,\alpha} {\phi,}_{\alpha}+ {\cal L}_{m}\right]
\end{equation}

where $\phi$ is the BD scalar field and $\omega$ is the BD
parameter. The matter content of the Universe is composed of
perfect fluid and viscous fluid given by
\begin{equation}
T_{\mu \nu}=(\rho+p+\pi)u_{\mu} u_{\nu}+(p+\pi)~g_{\mu \nu}
\end{equation}

where $u_{\mu}~u^{\nu}=-1$ and $\rho,~p$ are
respectively energy density and isotropic pressure of perfect fluid
and $\pi$ is the bulk viscous stress.\\

From the Lagrangian density $(1)$ we obtain the field equations as
\begin{equation}
G_{\mu \nu}=\frac{\omega}{{\phi}^{2}}\left[\phi  _{, \mu} \phi
_{, \nu} - \frac{1}{2}g_{\mu \nu} \phi _{, \alpha} \phi ^{,
\alpha} \right] +\frac{1}{\phi}\left[\phi  _{, \mu ; \nu} -g_{\mu
\nu}~ ^{\fbox{}}~ \phi \right]+\frac{1}{\phi}T_{\mu \nu}
\end{equation}
and
\begin{equation}
^{\fbox{}}~\phi=\frac{1}{3+2\omega}T
\end{equation}

where $T=T_{\mu \nu}g^{\mu \nu}$.\\

We now consider a homogeneous and anisotropic  space-time  model
described by the line  element

\begin{equation}
ds^{2}=-dt^{2}+a^{2}dx^{2}+b^{2}d\Omega_{k}^{2}
\end{equation}

where  $a$  and  $b$  are  functions  of  time  $t$ alone: we note
that

\begin{eqnarray}d\Omega_{k}^{2}= \left\{\begin{array}{lll}
dy^{2}+dz^{2}, ~~~~~~~~~~~~ \text{when} ~~~k=0 ~~~~ ( \text{Bianchi ~I ~model})\\
d\theta^{2}+sin^{2}\theta d\phi^{2}, ~~~~~ \text{when} ~~~k=+1~~
( \text{Kantowaski-Sachs~ model})\\
d\theta^{2}+sinh^{2}\theta d\phi^{2}, ~~~ \text{when} ~~~k=-1 ~~(
\text{Bianchi~ III~ model})\nonumber
\end{array}\right.
\end{eqnarray}

Here  $k$  is  the  curvature  index  of  the  corresponding
2-space, so  that  the  above  three  types  are  described  by
Thorne [19]  as  flat, closed  and  open respectively.\\

Now, in  BD  theory, the Einstein's  field  equations  for  the
above space-time  symmetry  are

\begin{equation}
\frac{\ddot{a}}{a}+2\frac{\ddot{b}}{b}=-\frac{1}{(3+2\omega)\phi}\left[
(2+\omega)\rho+3(1+\omega)(p+\pi)\right]-\omega\frac{\dot{\phi}^{2}}
{\phi^{2}}-\frac{\ddot{\phi}}{\phi}
\end{equation}

\begin{equation}
\frac{\dot{b}^{2}}{b^{2}}
+2\frac{\dot{a}}{a}\frac{\dot{b}}{b}=\frac{\rho}{\phi}-\frac{k}{b^{2}}-\left(\frac{\dot{a}}{a}+2\frac{\dot{b}}{b}
\right)\frac{\dot{\phi}}{\phi}+\frac{\omega}{2}\frac{\dot{\phi}^{2}}{\phi^{2}}
\
\end{equation}

and the wave equation for the BD scalar field $\phi$ is

\begin{equation}
\ddot{\phi}+\left(\frac{\dot{a}}{a}+2\frac{\dot{b}}{b}\right)\dot{\phi}=\frac{1}{3+2\omega}\left[\rho-3(p+\pi)\right]
\end{equation}

The energy conservation equation is
\begin{equation}
\dot{\rho}+\left(\frac{\dot{a}}{a}+2\frac{\dot{b}}{b}\right)(\rho+p)=0
\end{equation}

Here we consider the Universe to be filled with barotropic fluid
with EOS

\begin{equation}
p=\gamma\rho~,~~~0\le\gamma\le 1
\end{equation}

In full causal theory of non-equilibrium thermodynamics, $\pi$ is
given by the equation
\begin{equation}
\pi+\tau\dot{\pi}=-3\zeta
H-\frac{\varepsilon}{2}\tau\pi\left(3H+\frac{\dot{\tau}}{\tau}-\frac{\dot{\zeta}}{\zeta}-\frac{\dot{T}}{T}\right)
\end{equation}
where, $\zeta$ the coefficient of bulk viscosity,\quad $\tau$ is
the relaxation time for the bulk viscous effects, $T$ is the
temperature. Here, $H$ is the Hubble parameter given by
$H=\frac{\dot{S}}{S}=\frac{1}{3}(\frac{\dot{a}}{a}+2\frac{\dot{b}}{b})$,
where the average scale factor can be considered as
$S=\left(ab^{2}\right)^{1/3}$.\\

For $\tau=0$, (11) reduces to the non-causal equation $\pi=-3\zeta
H$. In (11), $\varepsilon=0$ gives the truncated theory while
$\varepsilon=1$ gives full Israel-Stewart-Hiscock theory [20]. It
is to be noted that in full Israel-Stewart-Hiscock theory, the
derivations from equilibrium are small, so we may assume $|\pi|\ll p$.\\

Now from the three equations (6) - (10), we see that there are
five equations and six unknown quantities, namely the scale
factors $a$ and $b$, the scalar field $\phi$, the density $\rho$
pressure $p$ and the bulk viscous stress $\pi$. So any one
quantity may be chosen freely to solve the system of equations.
Since the field equations contain $a$, $b$ and their derivatives,
so without any loss of generality, we shall assume the BD scalar
field $\phi$ is some power of the average scale factor, i.e.
$\phi = A\left(a b^{2}\right)^{\alpha}$ where $A$ and $ \alpha$
are constants. Substituting this value in the field equations, we
see that the field equations cannot be solved due to time
derivative occurs in the field equations. To solve the field
equations, at least in particular, let us consider the scale
factors in power law forms, given by $a=a_{0}t^{ m}$, $b=b_{0}t^{
n}$ where $m,~ n, ~a_{0},~b_{0}$ are positive constants.\\

So from the equations (6) - (10), we get

\begin{equation}
\beta+\frac{k}{b_{0}}t^{2(1-n)}=0
\end{equation}
where $\beta$ is given by

\begin{equation}
\beta=(m+2n)^{2}(-\frac\omega{2}\alpha^{2}-\omega\alpha)+(m+2n)(\omega\alpha+m-1)+3n^{2}
\end{equation}

From the above we see that for expanding Universe  the equation
(12) must be consistent for $n=1$ or $k=0$ (since $\beta$ is
constant).\\

For anisotropic Universe model the Hubble parameter and the
deceleration parameter are given by
$H=\frac{\dot{S}}{S}=\frac{m+2n}{3t}$  and
$q=-\frac{S\ddot{S}}{\dot{S}^{2}}=-1+\frac{3}{m+2n}$.\\

For accelerating Universe model, we have $q<0$. To support the
acceleration we find a relation between the powers of scale
factors given by $m+2n>3$. For $n=1$ we have $m>1$. But for $k =0$
we have $\beta=0$. In this case, $m+2n>3$ gives the domain of $n$
as $n>1+\sqrt{\frac{2+\omega\alpha+3\omega\alpha^{2}}{2}}$ or
$0<n<1-\sqrt{\frac{2+\omega\alpha+3\omega\alpha^{2}}{2}}$ with
$0<\alpha<0.52$ and $\omega>-3/2$.\\

From equations (7) and (8), we have the expressions of $\rho$ and
$\pi$ as

\begin{equation}
\rho=\rho_{0}t^{s}+\rho_{1}t^{r}
\end{equation}
\begin{equation}
\pi=\pi_{0}t^{s}+\pi_{1}t^{r}
\end{equation}
where, $r=(m+2n)\alpha-2n,~s=(m+2n)\alpha-2$ and $\rho_{0},\rho_{1},\pi_{0},\pi_{1}$ are constants.\\

Now we assume that the coefficient of bulk viscosity $\zeta$ and
the relaxation time $\tau$ are  simple power function of $\rho$.
Let, $\zeta=\zeta_{0}\rho^{\xi}$ and $\tau=\frac{\zeta}{\rho}$
where
$\xi$ and $\zeta_{0}(>0)$ are constants.\\

Now integrating equation (11) with the help of equations (14) and
(15), we have the expression for $T$ as

\begin{eqnarray*}
LogT= -\frac{2\rho_{1}^{1-\xi}}{\varepsilon\zeta_{0}(\xi
r-r-1)}~t^{1+r-\xi r }~_{2}F_{1}[\frac{r-\xi
r+1}{s-r},\xi-1,\frac{s-\xi
r+1}{s-r},-\frac{t^{s-r}\rho_{0}}{\rho_{1}}]-Log(\rho_{0}t^{s}+\rho_{1}t^{r})
\end{eqnarray*}
\begin{equation}
+\left\{\frac{2}{\varepsilon}+\frac{(\pi_{1}\rho_{0}-\pi_{0}\rho_{1})(s+2)}{(s-r)\pi_{0}\pi_{1}\alpha}\right\}Log(\pi_{0}t^{s}+\pi_{1}t^{r})
+\left(\frac{s+2}{\alpha}\right)\left\{1+\frac{(s\pi_{0}\rho_{1}-r\pi_{1}\rho_{0})}{(s-r)\pi_{0}\pi_{1}})\right\}Logt
\end{equation}

Now we consider two cases: $n=1$ and $k=0$.\\

{\bf Case I:} $n=1$: In this case, the expressions of $\rho$ and
$\pi$ can be written as

\begin{equation}
\rho=\rho_{2}t^{p_{1}}~~\text{and}~~\pi=\pi_{2}t^{p_{1}}
\end{equation}

where $p_{1}=(m+2)\alpha-2$, $\rho_{2}=\rho_{0}+\rho_{1}$ and
$\pi_{2}=\pi_{0}+\pi_{1}$.\\

So from equation (16), we get

\begin{equation}
T=T_{0}t^{A}e^{Bt^{p_{1}-p_{1}\xi+1}}
\end{equation}

where
$A=\frac{2p_{1}}{\varepsilon}+\frac{2(p_{1}+2)\rho_{2}}{\varepsilon\pi_{2}\alpha}+\frac{p_{1}+2}{\alpha}-p_{1}$,
$B=\frac{2}{\varepsilon\zeta_{0}\rho_{2}^{\xi-1}(p_{1}-p_{1}\xi+1)}$
and $T_{0}$ is a constant.\\

{\bf CaseII:} $k=0$: In this case, the expressions of $\rho$ and
$\pi$ become

\begin{equation}
\rho=\rho_{0}t^{s}~~\text{and}~~\pi=\pi_{0}t^{s}
\end{equation}

So from equation (16), we get

\begin{equation}
T=T_{0}t^{A_{1}}e^{B_{1}t^{s-\xi s+1}}
\end{equation}

where
$A_{1}=\frac{2s}{\varepsilon}+\frac{2(s+2)\rho_{0}}{\varepsilon\pi_{0}\alpha}+\frac{s+2}{\alpha}-s$
and $B_{1}=\frac{2}{\varepsilon\zeta_{0}\rho_{0}^{\xi-1}(s-\xi
s+1)}$.\\

\section{\normalsize\bf{discussions}}

We have considered an anisotropic space-time model in presence of
Brans-Dicke (BD) scalar field $\phi$, causal viscous fluid and
barotropic fluid. Here, we have found the exact solution for the
Bianchi-I, Kantowaski-Sachs and Bianchi-III models in a full
theory of non equilibrium thermodynamics. From the equations (18)
and (20) we see that the temperature $T$ is an explicit function
of time and these values clearly inconsistent for $\varepsilon=0$.
But for $\varepsilon=0$, equation (11) reduces to the truncated
theory. But $\varepsilon\ne 0$ support the full causal theory and
in this case the temperature $T$ is completely explicit function
of time $t$ for both the cases i.e. for $n=1$ and for $k=0$. For
$n=1$, we have shown that the acceleration is possible if $m>1$.
Also for $k=0$ i.e, for flat Universe, acceleration is possible if
$n$ satisfies some restriction, which is described before. If
$n<1$, the acceleration in the radial direction is not possible,
but in the transverse direction acceleration is possible as long
as $m+2n>3$ and effectively the anisotropic Universe will be
accelerate. Thus, in a full causal theory, the acceleration is
possible for anisotropic Universe. \\

{\bf Acknowledgement:}\\\\
One of the authors (UD) is thankful to UGC, Govt. of India for
providing research project grant (No. 32-157/2006(SR)).\\\\

{\bf References:}\\
\\
$[1]$ S. J. Perlmutter et al, {\it  Astrophys. J.} {\bf 517} 565
(1999); A. G. Rieses et al, {\it Astron. J.} {\bf 116} 1009
(1998); P. M. Garnavich et al,
{\it Astrophys. J.} {\bf 509} 74 (1998); G. Efstathiou et al, {\it astro-ph}/9812226. \\
$[2]$ B. Ratra and P. J. E. Peebles, {\it Phys. Rev. D} {\bf 37}
3406 (1988); R. R. Caldwell, R. Dave and P. J. Steinhardt, {\it
Phys. Rev. Lett.} {\bf 80} 1582 (1998).\\
$[3]$ S. Bridle, O. Lahav, J. P. Ostriker and P. J. Steinhardt,
{\it Science} {\bf 299} 1532 (2003); C. Bennett et al, {\it
Astrophys. J. Suppl.} {\bf 148} 1 (2003), {\it
astro-ph}/0302207; D. N. Spergel et al, {\it Astrophys. J. Suppl.} {\bf 148} 175 (2003), {\it astro-ph}/0302209.\\
$[4]$ V. Sahni and A. A. Starobinsky, {\it Int. J. Mod. Phys.}
{\bf 9} 373 (2003); T. Padmanabhan, {\it Phys. Rept.} {\bf 380} 235 (2003), {\it hep-th}/0212290.\\
$[5]$ N. Banerjee and D. Pavon, {\it Class. Quantum Grav.} {\bf 18} 593-599 (2001).\\
$[6]$ L. P. Chimento, A. S. Jakubi and D. Pavon, {\it Phys. Rev. D} {\bf 62} 063508 (2000).\\
$[7]$ C. Mathiazhagan and V. B. G. Johri, {\it Class. Quantum
Grav.} {\bf 1} L29 (1984); D. La and P. J. Steinhardt, {\it Phys. Rev. Lett.} {\bf 62} 376 (1989).\\
$[8]$ N. Banerjee and D. Pavon, {\it Phys. Rev. D} {\bf 63} 043504 (2001).\\
$[9]$ B. K. Sahoo and L. P. Singh, {\it Modern Phys. Lett. A} {\bf 18} 2725- 2734 (2003).\\
$[10]$ K. Nordtvedt, Jr., {\it Astrophys. J} {\bf 161} 1059
(1970); P. G. Bergmann, {\it Int. J. Phys.} {\bf 1} 25 (1968);
R. V. Wagoner, {\it Phys. Rev. D} {\bf 1} 3209 (1970).\\
$[11]$ O. Bertolami and P. J. Martins, {\it Phys. Rev. D} {\bf 61} 064007 (2000).\\
$[12]$ R. Maartens, {\it Class. Quantum Grav.} {\bf 12} 1455
(1995).\\
$[13]$ W. Zimdahl, {\it Phys. Rev. D} {\bf 53} 5483 (1996).\\
$[14]$ A. H. Guth, {\it Phys. Rev. D} {\bf 23} 347 (1981).\\
$[15]$ G. F. R. Ellis and M. S. Madsen, {\it Class. Quantum
Grav.} {\bf 8} 667 (1991).\\
$[16]$ W. A. Hiscock and L. Lindblom, {\it Ann. Phys.} (N. Y.)
{\bf 151} 466 (1983).\\
$[17]$ L. Lindblom, {\it Ann. Phys.} (N. Y.) {\bf 247} 1 (1996).\\
$[18]$ N. Banerjee and A. Beesham, {\it Aust. J. Phys.} {\bf 49}
899 (1996).\\
$[19]$  K. S. Thorne,  {\it Astrophys.  J.} {\bf 148} 51 (1967).\\
$[20v]$ R. Maartens, {\it astro-ph}/9609119.\\

\end{document}